\documentstyle[aps,pre,preprint]{revtex}
\tightenlines
\def\vperp{{\bf v}_\perp}
\def\vpar{{v}_\parallel}
\def\aperp{{\bf a}_\perp}
\def\apar{{a}_\parallel}
\def\la{\lesssim}
\def\ga{\gtrsim}

\begin{document}
\draft

\title{Apparent Deviations from Keplerian Acceleration for Stars Around the
Supermassive Black Hole at the Galactic Center}

\author{Abraham Loeb}

\address{Astronomy Department, Harvard University, 60 Garden St.,
Cambridge, MA 02138}

\date{\today}
\maketitle

\begin{abstract} 
We show that the time-dependent Doppler effect should induce measureable
deviations of the time history of the projected orbit of a star around the
supermassive black hole in the Galactic center (SgrA*) from the expected
Keplerian history.  In particular, the line-of-sight acceleration of the
star generates apparent acceleration of its image along its velocity vector
on the sky, even if its actual Keplerian acceleration in this direction
vanishes.  The excess apparent acceleration simply results from the
transformation of time between the reference frames of the observer and the
star.  Although the excess acceleration averages to zero over a full closed
orbit, it could lead to systematic offsets of a few percent in estimates of
the dynamical mass or position of the black hole that rely on partially
sampled orbits with pericentric distances of $\sim 10^{15}$ cm.  Deviations
of this magnitude from apparent Keplerian dynamics of known stars should be
detectable by future observations.

\end{abstract}


\section{Introduction}
The latest infrared snapshots of the inner arcsecond of the Milky Way
galaxy show bright stars orbiting around a common center of mass of $\sim
4 \times 10^6M_\odot$, providing unprecedented evidence for the existence
of a supermassive black hole at the location of the radio/infrared source,
SgrA* \cite{Ghez1,Genzel,Schodel}.  The inferred stellar orbits appear to
be Keplerian to within the current measurement errors. Nevertheless, one
would expect to see deviations from classical Keplerian orbits due to
various general-relativistic effects, that can in turn be used to constrain
the mass and spin of the black hole \cite{Jaroszynski,Pfahl}. Here we point
out that the leading-order correction to the apparent orbit of a star
involves a time transformation rather than a physical modification of the
orbit, and is of order $\sim \vert {\bf v}/c\vert$, where ${\bf v}$ is the
star velocity and $c$ is the speed of light.  Since some stars approach
pericenter at a few percent of the speed of light, future observations
reaching the percent level of sensitivity should detect this effect.  As we
show in \S 2, the naive Keplerian fit to the apparent orbit of the star in
these future data sets should reveal an acceleration residual due
to the time-dependent Doppler effect.  Although this excess acceleration
component averages to zero over a full closed orbit (during which the star
returns to its original position and velocity), it could lead to systematic
inconsistencies in estimates of the dynamical mass and position of the
black hole that are based on partially sampled orbits.

\section{Acceleration Residual Due to the Time-Dependent Doppler Effect} 
The standard Doppler transformation of infinitesimal time intervals between
the reference frames of the observer ($dt_{\rm obs}$) and the star ($dt$)
is given to leading-order in $\vert {\bf v}/c \vert$ by \cite{RL},
\begin{equation}
dt_{\rm obs}=dt \left(1 + {\vpar\over c}\right) ,
\label{eq:Doppler}
\end{equation}
where $\vpar (t)$ is the velocity of the star along the line-of-sight. The
observed transverse velocity of the star is therefore
different\footnote{The spatial line traced by the orbit remains unchanged
in the two reference frames. We ignore the Lorentz transformation of the
spatial coordinates since it corresponds to corrections of order $\sim
(v/c)^2$ or higher.} from its actual transverse velocity, $\vperp = ({d
{\bf x}_\perp/ dt})$,
\begin{equation}
{d {\bf x}_\perp\over dt_{\rm obs}}=
\left({dt\over dt_{\rm obs}}\right){d {\bf x}_\perp\over dt}\approx 
\left(1 - {\vpar \over c}\right) \vperp  ,
\label{eq:velocity}
\end{equation}
where the transverse position vector ${\bf x}_\perp$ corresponds to angular
coordinates on the sky times the distance to the Galactic center ($\simeq
8$ kpc \cite{Eisen}). Throughout the paper, the terms parallel or transverse
(perpendicular) are relative to the line-of-sight axis that starts at the
observer and goes through the star.

Taking the $t_{\rm obs}$--derivative of both sides of
equation~(\ref{eq:velocity}) and keeping terms to leading-order, we get two
Doppler components that contribute to the difference between the observed
and Keplerian values of the transverse acceleration of the star on the sky,
\begin{equation}
{\bf a}_{\perp, {\rm obs}}= \aperp - {2\vpar \over c} \aperp - {{\bf
v}_\perp\over c} \apar ,
\label{eq:acceleration}
\end{equation}
where ${\bf a}_{\rm obs}\equiv ({d^2 {\bf x}/dt^2_{\rm obs}})$ is the
observed acceleration and ${\bf a} \equiv ({d^2 {\bf x}/ dt^2})$ is the
actual Keplerian acceleration of the star.  These Doppler correction terms
were not included in past analysis of SgrA* data
\cite{Ghez1,Genzel,Schodel}.  The last term on the right-hand-side of
equation (\ref{eq:acceleration}) implies that the apparent transverse
acceleration gets a contribution from the Keplerian acceleration along the
line-of-sight, $\apar$. The image of the star may therefore show apparent
acceleration along its velocity vector on the plane of the sky even if its
Keplerian transverse acceleration vanishes, $\aperp =0$ (as long as
$\vperp\apar \neq 0$). Note that $\vperp$ is the transverse component of
the net relative velocity between the star and the observer, {\it
including} the transverse velocity of the Sun around the Galactic center.

A time-independent Doppler effect with $\apar=0$, such as expected for a
face-on orbit, gives ${\bf a}_{\perp, {\rm obs}}=(1-2 \vpar/c)\aperp$,
leading to a simple miscalibration of the black hole mass by a constant
factor of $(1-2\vpar/c)$. But more generally, the additional effect of the
last term in equation (\ref{eq:acceleration}) cannot be compensated for by
a simple fudge factor, as it depends on the particular parameters of the
stellar orbit and on the orbital phase.

The two excess acceleration terms on the right-hand-side of equation
(\ref{eq:acceleration}) amount to a fractional correction amplitude of a
few percent for the inferred value of $\aperp$ near pericenter of the
closest known stars to SgrA*. These stars reach their pericentric distance
of $\sim 100~{\rm AU}=1.5 \times 10^{15}~{\rm cm}$ with $(v/c)\la 3\times
10^{-2}$.  Omission of the above Doppler terms could lead to systematic
offsets in estimates of the black hole mass and position that are based on
partially sampled orbits.

Spectroscopic detection of the time-dependence of Doppler-shifted
frequencies, $\nu_{\rm obs}\propto (1-\vpar /c)$, of spectral lines emitted
from the atmosphere of the star (such as the Br $\gamma$ or He I features
reported for the star SO-2 in ref. \cite{Brgamma}) can be used to measure
$\vpar$ as a function of $t_{\rm obs}$ and thus infer the Keplerian
line-of-sight acceleration, $\apar = (1- \vpar/c)(d\vpar/dt_{\rm obs})$, as
a function of intrinsic time, $t=\int dt_{\rm obs} (1 - \vpar/c)$. This can,
in turn, be used to check for consistency with the astrometric derivation
of $\apar$ from equation (\ref{eq:acceleration}) using time-tagged
snapshots of the orbit of the star.  However, observers are likely to find
it much more challenging to obtain the required spectroscopic data than to
improve the precision on existing imaging techniques, for the purpose of
recovering $\apar$.

The leading-order Doppler effects in equation (\ref{eq:acceleration}) can
be easily incorporated into a computer program that searches for the
best-fit Keplerian orbit under the constraints of a given data set. The
{\it Doppler--corrected} Keplerian fit could provide direct constraints on
the radial motion of the star based on its apparent motion on the sky. Such
a fit would involve the same number of free parameters as in the standard
Keplerian fit.

\section{Discussion}
The time-dependent Doppler effect adds two {\it apparent} acceleration
terms to the Keplerian acceleration of a star on the sky in Equation
(\ref{eq:acceleration}). The last term on the right-hand-side of this
equation implies that the observer may see the star accelerating along the
projection of its velocity on the sky even if its Keplerian acceleration
vanishes in this direction. The existence of this term would be
particularly intriguing for a star on a highly eccentric (nearly radial)
orbit along the line-of-sight. The transverse motion of the sun around the
Galactic center with $({\bf v}_{\perp} /c)\approx 7 \times 10^{-4}$ would
induce a minimum transverse acceleration of the stellar image of $\sim
7\times 10^{-4} \apar$ in this case.

Additional special-relativistic or general-relativistic effects are of the
order of $\sim (v/c)^2$ or $\sim \phi/c^2$, or smaller, where $\phi$ is the
local gravitational potential probed by the star ($\phi \sim v^2$). These
corrections are at least an order of magnitude smaller than the Doppler
effect discussed here, for orbits with pericentric distances of $\ga
10^{15}$ cm.

Deviations from apparent Keplerian orbits may also be caused by
gravitational scattering off other stars or black hole companions to SgrA*
\cite{Rauch}. The time dependence of the excess acceleration caused by such
scattering events can be easily differentiated from that of the periodic
and fully-deterministic Doppler effect.

\bigskip
\bigskip
\paragraph*{Acknowledgments.}

The author thanks Andrea Ghez and Eric Pfahl for stimulating discussions.
This work was supported in part by NASA grants NAG 5-13292, and by
NSF grants  AST-0071019, AST-0204514.


\begin{references}

\bibitem{Ghez1} A. M. Ghez, {\it et al.}, Astrophys. J., submitted (2003),
astro-ph/0306130; A. M. Ghez, {\it et al.}, Astrophys. J. Lett., submitted
(2003), astro-ph/0309076

\bibitem{Genzel} R. Genzel, {\it et al.}, Astrophys.J. {\bf 594}, 812
(2003)

\bibitem{Schodel} R. Sch\"odel, {\it et al.}, Astrophys. J., submitted
(2003); astro-ph/0306214

\bibitem{Jaroszynski} M. Jaroszynski, Acta Astronomica, {\bf 48}, 653,
(1998)

\bibitem{Pfahl} E. Pfahl, \& A. Loeb, Astrophys. J., submitted (2003)
%ADD: ; astro-ph/

\bibitem{RL} G. B. Rybicki, \& A. P. Lightman, Radiative Processes
in Astrophysics, (New York: Wiley, 1979), p. 111

\bibitem{Eisen} F. Eisenhauer, {\it et al.}, Astrophys. J. Lett., submitted
(2003); astro-ph/0306220


\bibitem{Brgamma} A. M. Ghez, {\it et al.}, Astrophys. J., {\bf 586}, 127
(2003)

\bibitem{Rauch} K. P. Rauch, \& S. Tremaine, New Astronomy, {\bf 1}, 149
(1996)

\end{references}
\end{document}